# Observations of the recently discovered dwarf nova 1RXS J053234.9+624755, during the 2005 March superoutburst

**Gary Poyner & Jeremy Shears**

*A report of the Variable Star Section (Director: R. D. Pickard)*

1RXS J053234.9+624755 is a recently discovered dwarf nova. We present CCD and visual photometry during March 2005 of the first ever observed superoutburst in this system.

## Discovery circumstances

Discovery details of this new dwarf nova in Camelopardalis are intriguing and the result of real detective work. The Austrian observer Klaus Bernhard discovered it whilst comparing archival X-ray sources from the ROSAT all-sky bright source catalogue with variable stars in the ROTSE1 database.[1] An uncatalogued variable at 05h32m32s.68 +62°47′54″.7 in the ROTSE1 list was coincident with the X-ray source 1RXS J053234.9+624755. The new star was informally given the name 'Bernhard 01' by a group of observers who began to monitor for further outbursts. Initial reports suggested an outburst period of approximately 150 days, with the possibility of more frequent outbursts. The variability was eventually confirmed by Thomas Berthold (Sonneberg, Germany) who checked more than 200 plates and found six outbursts with a mean interval of 133.6d.[1,2] The star normally lies at mag 16, but reaches around magnitude 11 during outburst.

Although the data was limited, it was suggested that the star is a dwarf nova of the U Gem class.[1] The next outburst was thought to be due sometime in 2005 April. We didn't have to wait that long, however, as Wolfgang Kriebel (Germany) detected an outburst on March 16.809 at magnitude 12.0.[3]

## CCD photometry during the initial stages of outburst

Time resolved photometry was conducted on the evening of 2005 March 19, between 19.30 and 23.29 UT. The instrument was a Takahashi FS102 apochromatic refractor, aperture 0.1m, with an unfiltered Starlight Xpress MX716 CCD camera. The instrument is located at the Bunbury Observatory in Cheshire, UK, which is operated by one of the authors (JS). During the 4 hr observation window, 408 separate CCD images were collected, each with an exposure time of 30 seconds. Although the conditions were not ideal – bright moonlight and thin haze – the star was clearly visible in every image (Figure 1).

Each image was dark-subtracted and flat-fielded before being analysed photometrically using the 'multiple-images photometry' function of the AIP4WIN software.[4] No official chart with a comparison star sequence exists at the time of writing, but a provisional chart was kindly supplied by Mike Simonsen,[5] which has a sequence based on Tycho 2V and USNO A2.0v (Figure 2). The comparison star used in the photometry was '112' (mag 11.2) on this chart (also marked on Figure 1). The raw photometric data were subsequently imported into the *Peranso* software,[6] yielding the light curve shown in Figure 3. An inspection of the light curve shows approximately 3 complete sinusoidal cycles, with a maximum peak-to-trough amplitude of 0.3 mag (mag 11.7C to 12.0C – note C stands for unfiltered, i.e. clear, CCD). These appear to be classical 'superhumps' which are characteristic of the UGSU subclass of dwarf novae.

Statistical analysis of the data according to the ANOVA (**AN**alysis **O**f **VA**riance) method within *Peranso* indicates a superhump period of $0.0561 \pm 0.0015$d. Using *Peranso*, the light curve was folded at the measured period (Figure 4). This shows two cycles for clarity.

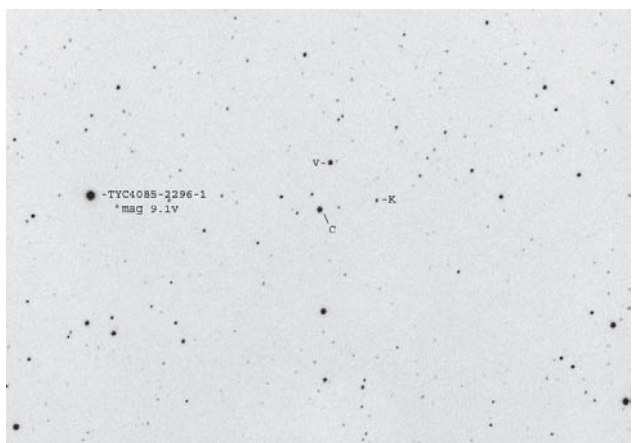

**Figure 1.** 1RXS J053234.9+624755 in outburst, 2005 March 19, 20.00 UT, Takahashi FS102, f/8. 1RXS J053234.9+624755 is marked 'V'. N is at the top. The comparison star 'C' was used in subsequent photometry and is the mag 11.2 star marked on a provisional AAVSO chart supplied by Mike Simonsen (Figure 2). *(J. Shears)*

Tonny Vanmunster (CBA Belgium) has reported the presence of superhumps during observations he made the previous night, 2005 March 18/19.[7] Hence our result is an independent confirmation. His analysis, also using the ANOVA method, yielded a superhump period of 0.0574 ± 0.0010d, which is consistent with our measurement.

mal outbursts and last up to 10 times longer. The light curve is characterised by superhumps, such as were detected in the outburst of 1RXS J053234.9+624755. These modulations are a few percent longer than the orbital period and are thought to be caused by precession of the accretion disc.[9]

## UGSU stars and superhumps

The classification and outburst mechanism of dwarf novae were discussed in two recent BAA *Journal* papers.[8,9] Collectively dwarf novae are referred to as U Gem (UG) stars after the prototype. There are three subclasses of UG stars: UGSS (after SS Cyg), UGZ (after Z Cam) and UGSU (after SU UMa). Dwarf novae are binary stars comprising a cool main sequence star or red dwarf (the secondary star) orbiting a white dwarf (primary). Matter flows from the secondary towards the white dwarf and forms an accretion disc around the white dwarf. From time-to-time the accretion disc flips between a dimmer, cooler state, to a hotter, brighter state, resulting in what we see as an outburst. A characteristic of the UGSU class of dwarf novae is that occasionally they exhibit 'superoutbursts', which are typically 0.5 to 1 magnitude brighter than nor-

## The later stages of outburst and the detection of flickering

The next clear night available for photometry was 2005 March 25. A time-resolved photometric run was conducted between 19.34 and 21.39 UT (terminated due to cloud cover). The star had faded to about 12.5C. This time the light curve was completely different (Figure 5): the smooth superhumps had been replaced by rapid modulations or flickering, with a mean amplitude of 0.12 mag. Period analysis using *Peranso* yielded no obvious period. However, we cannot rule out the possibility that the flickering masked an underlying superhump period.

Intense flickering was also detected visually on the evening of 2005 April 8 by GP. Initially a routine observation was made to check for the variable's brightness. It immediately became apparent that large amplitude, short-term variations were taking place, and it was a full five minutes before GP decided to attempt to record the phenomenon. The variations were so rapid that an attempt was made to estimate the magnitude every ten seconds for 24 minutes from 21.38 UT to 22.02 UT (Figure 6). By that time observer fatigue was becoming pronounced and monitoring was discontinued. The largest variation was 0.9 magnitude in twenty seconds. Obviously the accuracy expected from an observation like this will be less than the usual 0.1 magnitude which visual observers aim for. The time between estimates was very short (which was necessary as the variations were happening so quickly). Furthermore, just one comparison star was used (mag 14.7). However the light curve does show the high rate at which the star was varying.

It is clear that the extreme amplitude of the flickering had increased substantially from March 25 to April 8 (0.12 to 0.9 mag) as the outburst progressed. The cause of this phenomenon is surely worth further investigation during future outbursts – if indeed the flickering episode is repeated.

Further observations on the evenings of 2005 April 10 and 11 revealed further flickering, but to a lesser degree, whilst the star was at an average magnitude of 14.9 and returning to quiescence.

Flickering, or apparently random variations in brightness, is common during the outburst of many dwarf novae.[10] It can occur on many timescales, from a few seconds to hours. While the precise cause or causes of flickering are not fully understood, it may be because transfer of material from the secondary star to the white dwarf is a turbulent process, not a smooth one. Studies on U Gem suggest that flickering originates from the bright spot where material streaming from the secondary hits the accretion disc in a turbulent manner. By contrast other studies, notably of HT Cas, suggest that the flickering appears to arise predominantly in the turbulent inner regions of the accretion disc.[10] Whatever the cause, what is remarkable about the outburst of 1RXS J053234.9+624755 was the intensity of the flickering, which in the recent outburst was obvious to the visual observer.

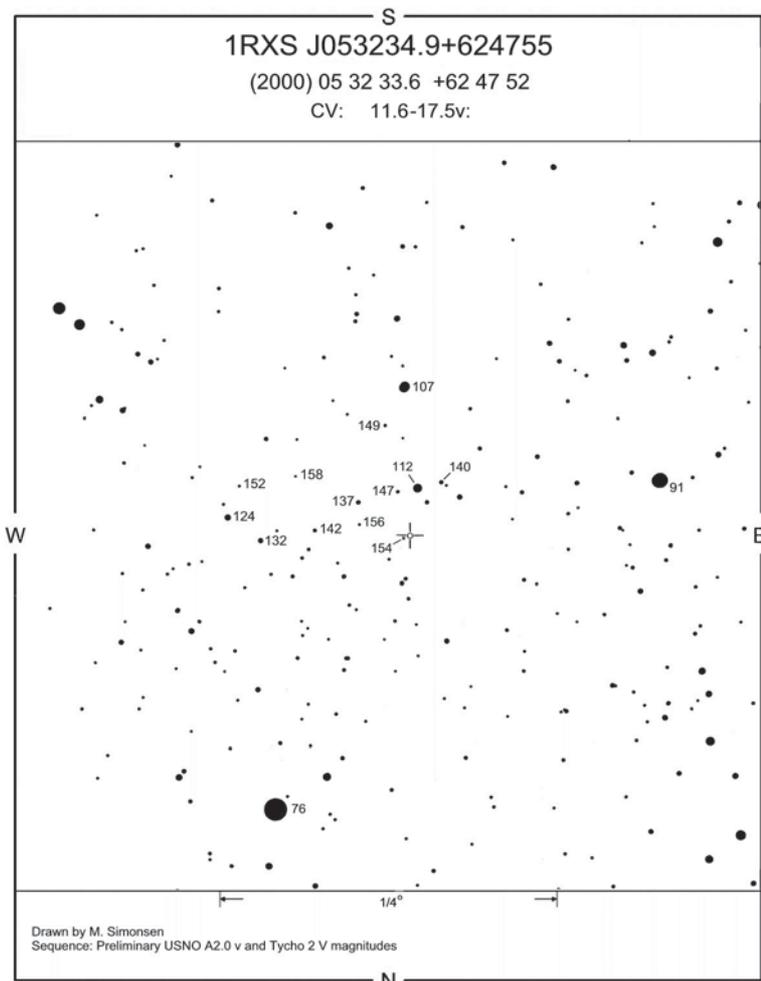

**Figure 2.** Provisional AAVSO chart for 1RXS J053234.9+624755.

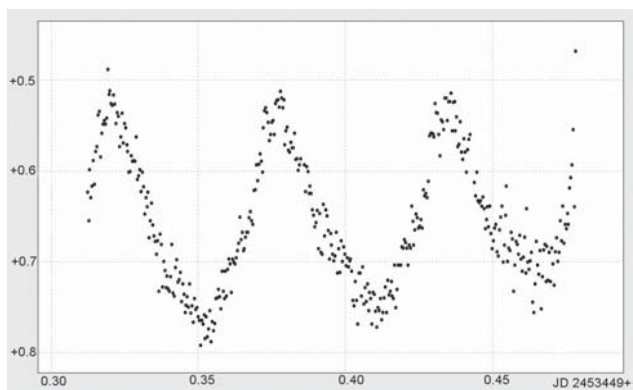

**Figure 3.** Superhumps in 1RXS J053234.9+624755 on 2005 March 19, 19.30 to 23.29 UT (magnitude relative to the mag 11.2 star on the provisional chart in Figure 2). *(J. Shears)*

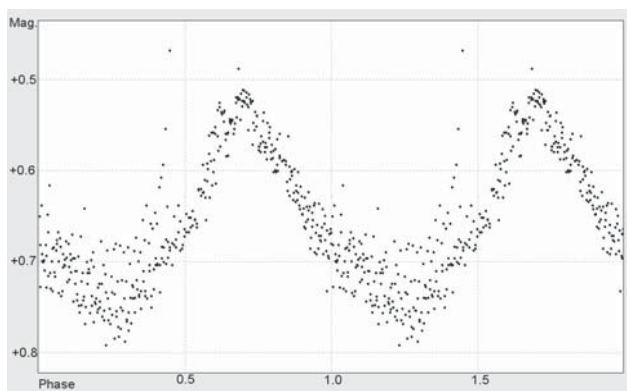

**Figure 4.** Phase diagram based on a superhump period of 0.0561d.

## Conclusion

Superhumps have been identified during the 2005 March outburst of 1RXS J053234.9+624755, confirming that this recently discovered dwarf nova is a member of the UGSU class. A superhump period of $0.0561 \pm 0.0015$d is consistent with a value obtained by Vanmunster 24 hours earlier. The later stages of the outburst were characterised by rapid, high amplitude flickering, which appeared to have replaced the superhumps. Given that this is the first outburst of this star which has been monitored by time-resolved photometry, follow-up measurements during future outbursts will be important in refining the nature of the superhumps and understanding the flickering. In addition, further monitoring to detect future out-bursts will be important in confirming the outburst frequency. 1RXS J053234.9+624755 is circumpolar from the UK, hence follow-up observations are encouraged.

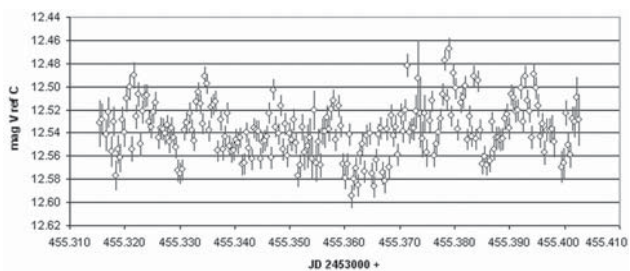

**Figure 5.** Photometry on 2005 March 25, 19.34 to 21.39 UT. Each point represents a 30sec exposure. *(J. Shears)*

## Acknowledgments

The authors would like to thank Mike Simonsen for giving permission to include his provisional chart for 1RXS J053234.9+624755, and Roger Pickard and David Boyd for their helpful comments during the preparation of this paper.

**Addresses:** **GP**: 67 Ellerton Road, Kingstanding, Birmingham, B44 0QE. [garypoyner@blueyonder.co.uk]
**JS**: 'Pemberton', School Lane, Bunbury, Tarporley, Cheshire, CW6 9NR. [bunburyobservatory@hotmail.com]

## References

1 Bernhard K. *et al.*, 'A new bright U Gem variable identified with the X-ray source 1RXS J053234.9+624755', *Information Bulletin on Variable Stars,* 5620 (2005)
2 CVnet-discussion group, http://groups.yahoo.com/group/cvnet-discussion
3 CVnet-outburst group, http://groups.yahoo.com/group/cvnet-outburst
4 Berry R. & Burnell J., *The Handbook of Astronomical Image Processing,* Willman–Bell, 2000
5 *Personal communication*. Provisional AAVSO chart used with the permission of Mike Simonsen.
6 Vanmunster T., *Peranso,* http://www.peranso.com
7 Vanmunster T., 'Detection of superhumps in the CV 1RXS J053234.9+624755', http://users.skynet.be/fa079980/cv_2005/1RXSJ053234_2005_mar_18.htm
8 Worraker W. J. & James N. D., 'Eclipsing dwarf novae', *J. Brit. Astron. Assoc.* **113**(2), 84 (2003)
9 Boyd D., 'Observations of the superoutburst of BC UMa in February 2003', *J. Brit. Astron. Assoc.* **113**(4), 208 (2003)
10 Hellier C,. *Cataclysmic Variable Stars: How and why they vary,* Springer–Verlag, 2001, Ch.10

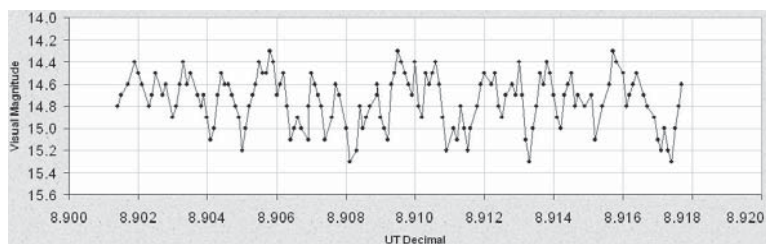

**Figure 6.** Visual light curve on 2005 April 8, 21.38 to 22.02 UT, showing flickering. *(G. Poyner)*